\documentclass[conference]{IEEEtran}
\IEEEoverridecommandlockouts
\usepackage{cite}
\usepackage{overpic}
\usepackage{amsmath,amssymb,amsfonts}
\usepackage{graphicx}
\usepackage{textcomp}
\usepackage{xcolor}
\usepackage{float}
\usepackage{amsthm}
\usepackage{graphicx}
\usepackage{epstopdf}
\usepackage{amsmath,bm}
\usepackage{amsfonts}
\usepackage{amssymb}
\usepackage{color}
\usepackage{multirow}
\usepackage{multicol}
\usepackage{soul,xcolor}
\usepackage{algorithm}
\usepackage{algpseudocode}%

\newcommand{\mathacr}[1]{\mathsf{#1}}
\theoremstyle{plain}

\newtheorem{lemma}{Lemma}

\newcommand{\vect}[1]{\mathbf{#1}}

\def\diag{\mathrm{diag}}

\def\kron{\otimes}
\def\tr{\mathrm{tr}}
\def\rank{\mathrm{rank}}
\def\Htran{\mbox{\tiny $\mathrm{H}$}}
\def\Ttran{\mbox{\tiny $\mathrm{T}$}}
\def\CN{\mathcal{N}_{\mathbb{C}}} 
\def\imagunit{\mathsf{j}} 

\def\sinc{\mathrm{sinc}}

\begin{document}

\title{Exploiting Array Geometry for Reduced-Subspace Channel Estimation in RIS-Aided Communications \vspace{-0.2cm}
\thanks{This work was supported by the  FFL18-0277 grant from the Swedish Foundation for Strategic Research and the Italian Ministry of Education and Research in the framework of the CrossLab project.}}

\author{\IEEEauthorblockN{\"Ozlem Tu\u{g}fe Demir$^*$, Emil Bj{\"o}rnson$^*$, Luca Sanguinetti$^{\dagger}$}
\IEEEauthorblockA{{$^*$Department of Computer Science, KTH Royal Institute of Technology, Kista, Sweden
		} \\ {$^\dagger$Dipartimento di Ingegneria dell'Informazione, University of Pisa, Pisa, Italy  
		} \\
		{Email: ozlemtd@kth.se, emilbjo@kth.se, luca.sanguinetti@unipi.it \vspace{-0.4cm}}
}
}

\maketitle

\begin{abstract}
A reconfigurable intelligent surface (RIS) can be used to improve the channel gain between a base station (BS) and user equipment (UE), but only if its $N$ reflecting elements are configured properly. This requires accurate estimation of the cascaded channel from the UE to the BS through each RIS element. If the channel structure is not exploited, pilot sequences of length $N$ must be used, which is a major practical challenge since $N$ is typically at the order of hundreds. To address this problem without requiring user-specific channel statistics, we propose a novel  estimator, called \emph{reduced-subspace least squares (RS-LS)} estimator, that only uses knowledge of the array geometry. The RIS phase-shift pattern is optimized to minimize the mean-square error of the channel estimates. The RS-LS estimator largely outperforms the conventional least-squares estimator,
and can be utilized with a much shorter pilot length since it exploits the fact that the array geometry confines the possible channel realizations to a reduced-rank subspace.
\end{abstract}


\section{Introduction}

Reconfigurable intelligent surface (RIS)-aided communication is one of the key areas explored for the next-generation wireless systems \cite{Wu2019,RISchannelEstimation_nested_knownBSRIS2,pei2021ris}. An RIS is a planar array of $N$ reflecting elements (meta-atoms) with sub-wavelength spacing. Each element can be configured by adjusting its impedance to induce a controllable phase-shift to the incident wave before it is reflected.
By optimizing the phase-shift pattern across the RIS, the reflected wavefront can be shaped (e.g., as a beam towards the intended receiver). To control each element based on its unique propagation path, we need to estimate the related channel coefficients. This is a key challenge since RISs are envisaged to consist of hundreds of elements \cite{RIS_emil_magazine}.

Conventionally, in RIS-aided communication between a base station (BS) and a user equipment (UE), the minimum pilot length for channel estimation is equal to $N$ (neglecting the uncontrollable direct BS-UE channel). Several methods that exploit sparsity, spatial channel correlation, and/or other specific characteristics of the channel can reduce the required pilot training length \cite{ris_training}. In \cite{ris_channel_estimation_lmmse, RISchannelEstimation_nested_knownBSRIS2,ris_joint_training_phase}, the spatial correlation among the channel coefficients is exploited to minimize either the mean-square error (MSE), or effective noise variance of the linear minimum MSE (LMMSE) estimator. However, the proposed methods are only applicable when the pilot length is at least $N$ (in the absence of a direct BS-UE channel). In addition, the LMMSE estimator requires knowledge of the complete high-dimensional spatial correlation matrices. These statistics are rather demanding to acquire in cases with a large number of BS antennas and/or RIS elements, and when the transmission consists of small data packets. To alleviate such difficulties, an alternative approach is to use the least squares (LS) estimator, which does not require any channel statistics. However, it requires the pilot length to be at least $N$.

Building on our recent paper \cite{Demir2021RISb} on multi-antenna communications without an RIS, we propose a novel channel estimator for RIS-aided communications that exploits the reduced-rank subspace created by the array geometry. This method improves the estimation quality without requiring user-specific statistical knowledge. The proposed \emph{reduced-subspace least squares (RS-LS)} estimator  outperforms the LS estimator and also enables shorter pilot lengths than $N$. We also derive the ideal RIS configuration pattern in the training phase that minimizes the MSE with the RS-LS estimator.
Numerical results are used to show that the projection of the ideal configurations to the closest unit-modulus RIS configurations provide significantly better performance than the benchmarks.

\vspace{-2mm}

\section{System and Channel Modeling}

We consider the RIS-assisted communication from an $M$-antenna BS to a single-antenna UE. The BS antennas are deployed as a uniform planar array (UPA) with $M_{\rm H}$ and $M_{\rm V}$ number of elements per row and per column, and we have $M=M_{\rm H}M_{\rm V}$ antennas in total. 
The RIS has $N$ reconfigurable elements, which form a UPA with $N_{\rm H}$ and $N_{\rm V}$ number of elements per row and per column, so that $N=N_{\rm H}N_{\rm V}$.  
Each RIS element is passive and introduces a phase-shift to the signals that impinge on it before reflection.

We consider a time-varying narrowband channel. Adopting the conventional block fading model, the time resources are divided into coherence blocks with static channel realizations \cite{massivemimobook}. We let $\tau_p$ denote the total number of samples allocated to pilot transmission per block. The channel from the UE to the RIS array is denoted by $\vect{h}\in \mathbb{C}^N$. In a planar array, there is always spatial correlation between the elements \cite{Bjornson2021b,Demir2021RISb}. To account for this, we adopt the spatially correlated Rayleigh fading model so that $\vect{h}\sim\CN(\vect{0}_N,\vect{R}_{\rm h})$ and it takes an independent realization in each coherence block. The spatial correlation matrix $\vect{R}_{\rm h}$ can generally be computed as
\begin{equation} \label{eq:spatial-correlation}
\vect{R}_{\rm h} = \beta_{\rm h}  \iint_{-\pi/2}^{\pi/2} f_{\rm h}(\varphi,\theta) \vect{a}(\varphi,\theta) \vect{a}^{\Htran}(\varphi,\theta) d\theta d\varphi 
\end{equation}
where $\beta_{\rm h}\geq0$ is the channel gain, $\varphi$ and $\theta$ are the azimuth and elevation angles. Here, $\vect{a}(\varphi,\theta)\in \mathbb{C}^N$ denotes the array response vector and $f_{\rm h}(\varphi,\theta)$ is the normalized spatial scattering function. We assume the RIS is deployed along the $y$ and $z$ axis and the waves only arrive from directions in front of it. Therefore, we have that $\varphi, \theta \in[-\frac{\pi}{2},\frac{\pi}{2}]$. The rank and subspaces of $\vect{R}_{\rm h}$ are determined by the array geometry (via $\vect{a}(\varphi,\theta)$) but also by the non-isotropic scattering and directivity pattern of the elements (via $f_{\rm h}(\varphi,\theta)$) \cite{Demir2021RISb, Demir2021RIS}. In an \emph{isotropic scattering environment} where the multipath components are equally strong in all directions and the antennas are isotropic, the spatial scattering function is given by $f_{\rm h}(\varphi,\theta) = \cos(\theta) / (2\pi)$.\footnote{The cosine of the elevation angle, $\cos(\theta)$, comes from the differential of the solid angle in the spherical coordinate system.} We denote the resulting normalized correlation matrix by $\vect{R}_{\rm iso}$ and the $(m,l)$th entry is~\cite{Bjornson2021b}
\begin{equation}\label{R-iso}
    \left[\vect{R}_{\rm iso} \right]_{m,l} =  \sinc \left( 2 \sqrt{\left(d_{\rm H}^{ml}\right)^2+\left(d_{\rm V}^{ml}\right)^2}\right)
\end{equation}
where $\sinc(x) = \sin(\pi x)/ (\pi x)$ is the sinc function. The horizontal and vertical distances (normalized by the wavelength) between antenna (or RIS element) $m$ and $l$ are denoted by $d_{\rm H}^{ml}$ and $d_{\rm V}^{ml}$, respectively. 

Let $g_{m,n}\in \mathbb{C}$ denote the channel from RIS element $n$ to BS antenna $m$, for $n=1,\ldots,N$ and $m=1,\ldots,M$. We let $\vect{g}_m=[g_{m,1} \ \ldots \ g_{m,N}]^{\Ttran}\in \mathbb{C}^{N}$ be the vector collecting the channels from the RIS to BS antenna $m$ and $\vect{g}^{\prime}_n=[g_{1,n} \ \ldots \ g_{M,n}]^{\Ttran}\in \mathbb{C}^{M}$ the vector collecting channels from RIS element $n$ to the BS. Using the Kronecker model \cite{Shiu2000a,Demir2021RIS} with the receive and transmit correlation matrices for the BS and RIS, $\vect{R}_{\mathrm{g}^{\prime}}\in \mathbb{C}^{M\times M}$ and $\vect{R}_{\mathrm{g}}\in \mathbb{C}^{N \times N}$, respectively, the channel vectors are distributed as $\vect{g}_m\sim\CN(\vect{0}_N,[\vect{R}_{\mathrm{g}^{\prime}} ]_{m,m}\vect{R}_{\mathrm{g}})$ and  $\vect{g}^{\prime}_n\sim\CN(\vect{0}_M,[\vect{R}_{\mathrm{g}} ]_{n,n}\vect{R}_{\mathrm{g}^{\prime}})$. We assume $\vect{h}$ and $\vect{g}_m$ are independent.

\section{Pilot Transmission and Channel Estimation}

In this paper, we assume the direct link between the BS and UE is negligible to focus on the RIS phase-shift design during pilot transmission. Note that no generality is lost by this assumption since the direct channel can be estimated separately by an  orthogonal pilot scheme \cite{RISchannelEstimation_nested_knownBSRIS2,Demir2021RIS}. 
To optimize the RIS phase-shifts and signal precoding during data transmission, in each coherence block, the BS must estimate the cascaded channel 
\begin{equation}
\vect{h}_m\triangleq \vect{h}\odot\vect{g}_m \in \mathbb{C}^N
\end{equation}
for $m=1,\ldots,M$, where $\odot$ denotes the Hadamard product. The UE sends a predefined pilot sequence $\boldsymbol{\psi}\in \mathbb{C}^{\tau_p}$ during $\tau_p$ channel uses and the RIS switches its configuration between each channel use to explore different channel dimensions.
 The received signal at BS antenna $m$ during this training phase is 
\begin{equation}
    \vect{y}_m = \sqrt{\rho}\diag(\boldsymbol{\psi})\vect{\Phi}\vect{h}_m + \vect{n}_m, \quad m=1,\ldots,M,
\end{equation}
where $\diag(\boldsymbol{\psi}) \in \mathbb{C}^{\tau_p \times \tau_p}$ is the diagonal matrix with entries from $\boldsymbol{\psi}$, and $\vect{n}_m\sim \CN\left(\vect{0}_{\tau_p}, \vect{I}_{\tau_p} \right)$ is the thermal noise vector whose samples are independent between different BS antennas. Also, $\rho>0$ is the pilot signal-to-noise ratio (SNR) and $\vect{\Phi}\in \mathbb{C}^{\tau_p\times N}$ is the RIS phase-shift matrix with elements $\left[\vect{\Phi}\right]_{t,n}=e^{-\imagunit\phi_{t,n}}$ where $\phi_{t,n}$ represents the phase-shift used by RIS element $n$ at pilot time index $t$. For simplicity, we select $\boldsymbol{\psi}$ as the all-ones vector.\footnote{This selection does not lead to loss of generality since the cumulative phase-shift of the pilot signals and the RIS can be represented by the entries of the RIS phase-shift matrix $\vect{\Phi}$.} With this selection and collecting the received signals for all the BS antennas, we obtain
\begin{equation} \label{eq:concat}
   \underbrace{ \begin{bmatrix}\vect{y}_1 \\ \vdots \\ \vect{y}_M \end{bmatrix}}_{\triangleq \vect{y}}= \sqrt{\rho}\underbrace{\left(\vect{I}_M \kron \vect{\Phi}\right)}_{\triangleq \vect{\Phi}_M} \underbrace{ \begin{bmatrix}\vect{h}_1 \\ \vdots \\ \vect{h}_M \end{bmatrix}}_{\triangleq \vect{x}} + \ \underbrace{ \begin{bmatrix}\vect{n}_1 \\ \vdots \\ \vect{n}_M \end{bmatrix}}_{\triangleq \vect{n}}.
\end{equation}
We assume that the spatial correlation matrix $\vect{R}_{\rm x}$ of the channel vector ${\bf x}$ is \emph{not known} at the BS (due to the practical challenge of acquiring $M^2N^2$ entries). In these circumstances, ${\bf x}$ can be estimated by the LS estimator~\cite{Kay1993a}, but it requires $\tau_p = N$, which can be large. 
In the context of holographic massive MIMO channels, \cite{Demir2021RISb} proposed a more efficient estimator that only exploits knowledge of the UPA geometry.
Inspired by~\cite{Demir2021RISb}, we will now propose a RS-LS channel estimator for RIS-aided communications that can be applied for  $\tau_p<N$.

\subsection{Reduced-Subspace Least Squares Estimation}
 The spatial correlation matrix of $\vect{x}$ is given by $\vect{R}_{\rm x}=\vect{R}_{{\rm g}^{\prime}} \kron \left(\vect{R}_{\rm h}\odot\vect{R}_{\rm g}\right)$ from the  independence of the channels $\vect{h}$ and $\{\vect{g}_m\}$. Let $\vect{R}_{\rm x}=\vect{U}\vect{D}\vect{U}^{\Htran}$ be the eigendecomposition of  $\vect{R}_{\rm x}$ and denote the rank as $r=\rank(\vect{R}_{\rm x})$. The idea of RS-LS channel estimation is that the channel vector $\vect{x}$ can be expressed as $\vect{U}_1\vect{w}$ where the elements of $\vect{w}$ are independent and $\vect{U}_1\in \mathbb{C}^{MN\times r}$ is the matrix whose columns are the orthonormal eigenvectors of $\vect{R}_{\rm x}$ corresponding to its $r$ non-zero eigenvalues. The elements of $\vect{w}$ have different and \emph{unknown} variances.
We propose to obtain the RS-LS estimate of $\vect{x}$ as follows:
\begin{enumerate}
\item  Obtain the LS estimate of $\vect{w}$ in the subspace spanned by the columns of $\vect{U}_1$;
\item  Bring the estimate back to the original $MN$-dimensional space by multiplying the signal by $\vect{U}_1$. 
\end{enumerate}
Assuming $M\tau_p\geq r$, the RS-LS estimate of $\vect{x}$ takes the form
\begin{align} \label{eq:RS-LS-estimate}
    \widehat{\vect{x}}_{\rm RS-LS} = \frac{1}{{\sqrt{\rho}}}\vect{U}_1\left(\vect{U}_1^{\Htran}\vect{\Phi}_M^{\Htran}\vect{\Phi}_M\vect{U}_1\right)^{-1}\vect{U}_1^{\Htran}\vect{\Phi}_M^{\Htran}\vect{y}.
\end{align}
The RS-LS estimation method relaxes the original requirement $M\tau_p\geq MN$ for the LS estimation of $\vect{x}$ to $M\tau_p\geq r$ and removes noise from all unused channel dimensions when $r<MN$. Using Kronecker product identities, we can express $\vect{U}_1$ as $\vect{U}_1=\vect{U}_{{\rm BS},1}\kron\vect{U}_{{\rm RIS},1}$ where $\vect{U}_{{\rm BS},1}\in\mathbb{C}^{M\times r_{\rm BS}}$ and $\vect{U}_{{\rm RIS},1}\in\mathbb{C}^{N\times r_{\rm RIS}}$ consist of the orthonormal eigenvectors (corresponding to the non-zero eigenvalues) of the correlation matrices
$\vect{R}_{{\rm g}^{\prime}}$ and $\vect{R}_{\rm h}\odot\vect{R}_{\rm g}$, respectively. These matrices represent the spatial correlation characteristics at the BS and RIS side, respectively. We have $r_{\rm BS}=\rank(\vect{R}_{{\rm g}^{\prime}})$, $r_{\rm RIS}=\rank(\vect{R}_{\rm h}\odot\vect{R}_{\rm g})$, and $r=r_{\rm BS}r_{\rm RIS}$.

The RS-LS estimator defined above requires knowledge of the subspace spanned by $\vect{R}_{\rm x}$. However, we can alleviate this by instead considering the subspace spanned by another spatial correlation matrix $\overline{\vect{R}}_{\rm x}$ representing the union of the span of all \emph{plausible} correlation matrices. 
The following lemma proves that we can select $\overline{\vect{R}}_{\rm x}$ based on the isotropic scattering case.

\begin{lemma} \label{lemma:span} Let $\overline{\vect{R}}_{\rm x }=\overline{\vect{R}}_{{\rm g}^{\prime}} \kron \left(\overline{\vect{R}}_{\rm h}\odot\overline{\vect{R}}_{\rm g}\right)$ and $\vect{R}_{\rm x}=\vect{R}_{{\rm g}^{\prime}} \kron \left(\vect{R}_{\rm h}\odot\vect{R}_{\rm g}\right)$ be two spatial correlation matrices for the cascaded channel obtained using the same RIS and BS array geometry. The spatial scattering functions corresponding to the correlation matrices $\overline{\vect{R}}_{i}$ and $\vect{R}_i$, according to the model in \eqref{eq:spatial-correlation},  are denoted by $\overline{f}_{i}(\varphi,\theta)$ and $f_i(\varphi,\theta)$, respectively, for $i\in \{{\rm g}^{\prime},{\rm h}, {\rm g}\}$, $\varphi\in[-\pi/2,\pi/2]$ and $\theta\in[-\pi/2,\pi/2]$. Assume that the spatial scattering functions are either continuous at each point on its domain or contain Dirac delta functions.

If the domain of  $\overline{f}_i(\varphi,\theta)$ for which $\overline{f}_i(\varphi,\theta)>0$ contains the domain $f_i(\varphi,\theta)$ for which $f_i(\varphi,\theta)>0$, for $i\in\{{\rm g}^{\prime},{\rm h},{\rm g}\}$ then the subspace spanned by the columns of $\overline{\vect{R}}_{\rm x}$ contains the subspace spanned by the columns of $\vect{R}_{\rm x}$.
\begin{proof}
The proof follows similar steps as in \cite[Lem.~3]{Demir2021RISb} and is omitted due to the limited space. Apart from mathematical manipulations, 
the key new aspects are to define the new spatial scattering functions $f_{{\rm g}^{\prime}}(\varphi_1,\theta_1)f_{\rm h}(\varphi_2,\theta_2)f_{\rm g}(\varphi_3,\theta_3)$ and array response vectors $\vect{a}_{\rm BS}(\varphi_1,\theta_1)\kron\left(\vect{a}_{\rm RIS}(\varphi_2,\theta_2)\odot\vect{a}_{\rm RIS}(\varphi_3,\theta_3)\right)$ for the UPAs in terms of $(\varphi_1,\theta_1,\varphi_2,\theta_2,\varphi_3,\theta_3)$ on the six-dimensional angular domain.
\end{proof}
\end{lemma}

The spatial scattering function is non-zero for all angles in the isotropic scattering case.
Following Lemma~\ref{lemma:span}, when we use  $\overline{\vect{R}}_{\rm x}=\vect{R}_{\rm BS, iso}\kron\left(\vect{R}_{\rm RIS, iso}\odot\vect{R}_{\rm RIS, iso}\right)$, where $\vect{R}_{\rm BS, iso}$ and $\vect{R}_{\rm RIS, iso}$  are the spatial correlation matrices for isotropic scattering from \eqref{R-iso} (according to the BS or RIS array geometry, respectively) in the RS-LS estimator, 
we ensure that all \emph{plausible} channel subspaces are included. Hence, any $\vect{x}$ can be expressed as $\overline{\vect{U}}_1\overline{\vect{w}}$ for some reduced-dimension vector $\overline{\vect{w}} \in \mathbb{C}^{\overline{r}}$. 
The columns of $\overline{\vect{U}}_1\in \mathbb{C}^{MN \times \overline{r}}$ 
are the orthonormal eigenvectors corresponding to the  $ \overline{r}$ non-zero eigenvalues of $\overline{\vect{R}}_{\rm x}$. We call this the \emph{conservative RS-LS} estimator:  \vspace{-2mm}
\begin{align} \label{eq:RS-LS-estimate-approx}
    &\widehat{\vect{x}}_{\rm RS-LS}^{\rm conserv} = \frac{1}{\sqrt{\rho}}\overline{\vect{U}}_1\left(\overline{\vect{U}}_1^{\Htran}\vect{\Phi}_M^{\Htran}\vect{\Phi}_M\overline{\vect{U}}_1\right)^{-1}\overline{\vect{U}}_1^{\Htran}\vect{\Phi}_M^{\Htran}\vect{y} \nonumber\\
    &= \underbrace{\overline{\vect{U}}_1\overline{\vect{w}}}_{=\vect{x}}+\frac{1}{\sqrt{\rho}}\overline{\vect{U}}_1\left(\overline{\vect{U}}_1^{\Htran}\vect{\Phi}_M^{\Htran}\vect{\Phi}_M\overline{\vect{U}}_1\right)^{-1}\overline{\vect{U}}_1^{\Htran}\vect{\Phi}_M^{\Htran}\vect{n}
\end{align}
where we assume $\overline{r}\leq M \tau_p\leq M N$. The respective MSE is
\begin{align} \label{eq:mse-rsls}
    {\rm MSE}_{\rm RS-LS}^{\rm conserv}&= \frac{1}{{\rho}}\tr\left(\overline{\vect{U}}_1\left(\overline{\vect{U}}_1^{\Htran}\vect{\Phi}_M^{\Htran}\vect{\Phi}_M\overline{\vect{U}}_1\right)^{-1}\overline{\vect{U}}_1^{\Htran}\right)\nonumber \\
    &=\frac{1}{{\rho}}\tr\left(\left(\overline{\vect{U}}_1^{\Htran}\vect{\Phi}_M^{\Htran}\vect{\Phi}_M\overline{\vect{U}}_1\right)^{-1}\right)
\end{align}
and depends on the RIS phase-shift matrix $\vect{\Phi}$. In the next section, we will search for the $\vect{\Phi}$ that minimizes   ${\rm MSE}_{\rm RS-LS}^{\rm conserv}$.

 \section{Optimized RIS Configuration for Estimation}
 
Our aim is to find the phase-shift matrix $\vect{\Phi}$ that minimizes~\eqref{eq:mse-rsls} under the unit-modulus constraints
 $|\left[\vect{\Phi}\right]_{t,n}|=1$, for $t=1,\ldots,\tau_p$ and $n=1,\ldots,N$.  To obtain a closed-form solution, we first relax this problem as
\begin{align}
& \underset{\tr\left(\vect{\Phi}^{\Htran}\vect{\Phi}\right)\leq N\tau_p}{\mathacr{minimize}} \ \ \tr\left(\left(\overline{\vect{U}}_1^{\Htran}\vect{\Phi}_M^{\Htran}\vect{\Phi}_M\overline{\vect{U}}_1\right)^{-1}\right)  \label{eq:rsls-opt}
\end{align}
where the non-convex unit-modulus constraints are replaced by a Frobenius norm constraint on the matrix $\vect{\Phi}$. After obtaining the optimal solution to the relaxed problem in \eqref{eq:rsls-opt}, we will project it to guarantee $|\left[\vect{\Phi}\right]_{t,n}|=1$, for $t=1,\ldots,\tau_p$ and $n=1,\ldots,N$. 
Let the Kronecker decomposition $\overline{\vect{U}}_1=\overline{\vect{U}}_{{\rm BS},1}\kron\overline{\vect{U}}_{{\rm RIS},1}$ be constructed based on the eigenspaces of the BS and RIS spatial correlation matrices selected for the conservative RS-LS estimation. By using the fact that $\vect{\Phi}_M=\vect{I}_M \kron \vect{\Phi}$, the objective function in \eqref{eq:rsls-opt} is written as
\begin{align}
&\tr\Bigg(\bigg(\left(\overline{\vect{U}}_{{\rm BS},1}^{\Htran}\kron\overline{\vect{U}}_{{\rm RIS},1}^{\Htran}\right)\left(\vect{I}_M \kron \vect{\Phi}^{\Htran}\right)\left(\vect{I}_M \kron \vect{\Phi}\right)\nonumber\\
&\quad \times\left(\overline{\vect{U}}_{{\rm BS},1}\kron\overline{\vect{U}}_{{\rm RIS},1}\right)\bigg)^{-1}\Bigg) \nonumber\\
&= \tr\left(\left(\vect{I}_{\overline{r}_{\rm BS}}\kron \left(\overline{\vect{U}}_{{\rm RIS},1}^{\Htran}\vect{\Phi}^{\Htran}\vect{\Phi}\overline{\vect{U}}_{{\rm RIS},1}\right)\right)^{-1}\right) \nonumber \\
&=\overline{r}_{\rm BS}\times\tr\left(\left(\overline{\vect{U}}_{{\rm RIS},1}^{\Htran}\vect{\Phi}^{\Htran}\vect{\Phi}\overline{\vect{U}}_{{\rm RIS},1}\right)^{-1}\right)\label{eq:rsls-opt2}
\end{align}
where we have used the distributive properties of the Kronecker product. We have also implicitly assumed that $\vect{A}\triangleq\vect{\Phi}\overline{\vect{U}}_{{\rm RIS},1}\in \mathbb{C}^{\tau_p \times \overline{r}_{\rm RIS}}$ has full column rank, i.e., $\rank(\vect{A})=\overline{r}_{\rm RIS}\leq \tau_p$. Denote by $\vect{A}=\vect{S}_{\rm A}\vect{\Lambda}_{\rm A}\vect{V}_{\rm A}^{\Htran}$ the singular value decomposition of $\vect{A}$ with non-zero singular values $\lambda_{{\rm A},1}\geq \ldots \geq \lambda_{{\rm A},\overline{r}_{\rm RIS}}>0$. Then, we can rewrite~\eqref{eq:rsls-opt2} as 
\begin{align}
  \overline{r}_{\rm BS} \sum_{i=1}^{\overline{r}_{\rm RIS}} \frac{1}{\lambda_{{\rm A},i}^2} \label{eq:objectiveb}
\end{align}
whose minimum over the singular values is a monotonically decreasing function of $\sum_{i=1}^{\overline{r}_{\rm RIS}}\lambda_{{\rm A},i}^2=\tr(\vect{A}^{\Htran}\vect{A})$, which is equal to
\begin{align}
    &\tr\left(\overline{\vect{U}}_{{\rm RIS},1}\overline{\vect{U}}_{{\rm RIS},1}^{\Htran}\vect{\Phi}^{\Htran}\vect{\Phi}\right)\nonumber\\&\hspace{1mm}=  \tr\left(\overline{\vect{U}}_{{\rm RIS}}\overline{\vect{U}}_{{\rm RIS}}^{\Htran}\vect{\Phi}^{\Htran}\vect{\Phi}\right)- \tr\left(\overline{\vect{U}}_{{\rm RIS},2}\overline{\vect{U}}_{{\rm RIS},2}^{\Htran}\vect{\Phi}^{\Htran}\vect{\Phi}\right) \nonumber\\
    &=\tr\left(\vect{\Phi}^{\Htran}\vect{\Phi}\right)- \tr\left(\overline{\vect{U}}_{{\rm RIS},2}^{\Htran}\vect{\Phi}^{\Htran}\vect{\Phi}\overline{\vect{U}}_{{\rm RIS},2}\right) \leq N\tau_p
\end{align}
where $\overline{\vect{U}}_{{\rm RIS},2}\in \mathbb{C}^{N \times (N-\overline{r}_{\rm RIS})}$ is the matrix whose columns are the orthonormal eigenvectors of $\overline{\vect{R}}_{\rm h}\odot\overline{\vect{R}}_{\rm g}$ corresponding to the zero-valued eigenvalues. The above inequality is satisfied with equality when the right singular vectors of $\vect{\Phi}$ corresponding to non-zero singular values lie in the subspace spanned by $\overline{\vect{U}}_{{\rm RIS},1}$, i.e., $\vect{\Phi}\overline{\vect{U}}_{{\rm RIS},2}=\vect{0}_{\tau_p \times (N-\overline{r}_{\rm RIS})}$. In this way, the objective value in \eqref{eq:objectiveb} is minimized. Moreover, $\lambda_{{\rm A},i}= \sqrt{N\tau_p/\overline{r}_{\rm RIS}}$, for $i=1,\ldots,\overline{r}_{\rm RIS}$ to minimize \eqref{eq:objectiveb}. We can construct the optimal $\vect{\Phi}$ that satisfies all these constraints as 
\begin{align} \label{eq:optimal-phase-shift}
    \vect{\Phi}^{\star} = \sqrt{\frac{N\tau_p}{\overline{r}_{\rm RIS}}}\vect{S}_{{\rm \Phi},1}\overline{\vect{U}}_{{\rm RIS},1}^{\Htran}
\end{align}
where $\vect{S}_{{\rm \Phi},1}\in \mathbb{C}^{\tau_p\times \overline{r}_{\rm RIS}}$ is an arbitrary matrix with orthonormal columns. Lastly, a unit-modulus phase-shift matrix is obtained as $\vect{\Phi}=e^{\imagunit\angle{ \vect{\Phi}^{\star}}}$, where we only keep the phase-shifts.

\section{Numerical Results}

Numerical results are now used to quantify the performance of the proposed channel estimation method and phase-shift design, and to compare them with benchmarks, in terms of the normalized MSE (NMSE).
We assume the BS is equipped with $M=64$ antennas, and deployed as a square UPA with $M_{\rm H}=M_{\rm V}=8$. The horizontal and vertical inter-antenna distances are $\lambda/4$, where $\lambda$ is the wavelength. The RIS is equipped with $N=256$ elements, which are deployed as a square UPA with $N_{\rm H}=N_{\rm V}=16$. The horizontal and vertical inter-element distances are $\lambda/8$. The spatial correlation matrices are generated as in~\cite[Sec.~IV]{Demir2021RISb} that follows the clustered scattering model with the exponential power delay profile from \cite[p.~54-58]{series2017guidelines}.  For the proposed conservative RS-LS estimation, the spatial matrices for isotropic scattering, namely $\vect{R}_{\rm BS, iso}$ and $\vect{R}_{\rm RIS, iso}$, are computed according to~\eqref{R-iso}. 

\begin{figure}[t!]
\hspace{0mm}
\includegraphics[trim={0.7cm 0.1cm 1.5cm 1cm},clip,width=3.2in]{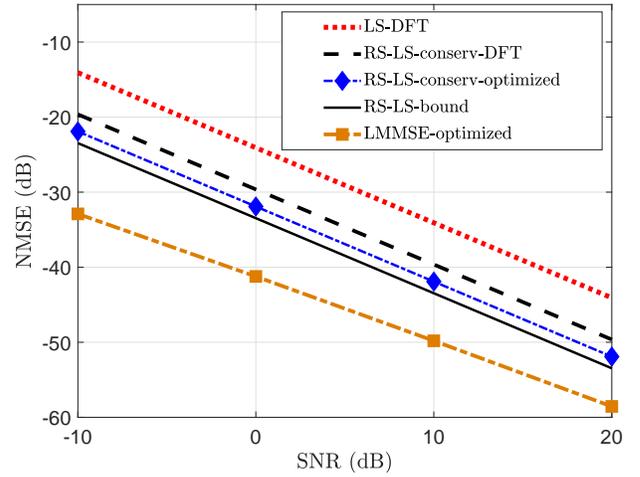}
			\vspace{-0.1cm}
			\caption{NMSE versus SNR for different estimators and RIS phase-shift configurations with $\tau_p=N=256$.} \label{fig:RSLS-taup-N} \vspace{-2mm}
\end{figure}

In Fig.~\ref{fig:RSLS-taup-N}, the pilot length $\tau_p$ is set to $\tau_p=N$. The ``bound'' in the legends represents the NMSE when we do not project the optimal RIS phase-shift matrix to have unit-modulus entries. The ``optimized'' scheme corresponds to the proposed conservative RS-LS estimator in \eqref{eq:RS-LS-estimate-approx} when we use the projected unit-modulus RIS phase-shift matrix from the optimal $\vect{\Phi}^{\star}$. In addition, we consider three benchmarks: i) the conventional LS estimator; ii) the conservative RS-LS estimator with the DFT matrix as the RIS phase-shift matrix; and iii) the LMMSE estimator with the same phase-shift matrix as the proposed ``optimized'' scheme.\footnote{The best phase-shift configuration is adopted for the benchmark channel estimators (LMMSE and LS) among the considered alternatives.}  The ``DFT'' case corresponds to one of the alternative optimal phase-shift matrices when $\tau_p=N=\overline{r}_{\rm RIS}$ from \eqref{eq:optimal-phase-shift}. As expected, the LMMSE estimator results in the lowest NMSE since it exploits the full spatial correlation matrix. On the other hand, the proposed conservative RS-LS method reduces the NMSE significantly compared to the conventional LS estimator by exploiting the spatial correlation that is induced by the array geometries of BS and RIS. Note that this is all the information the conservative RS-LS method needs to operate, while the LMMSE estimator requires the full $MN \times MN$ correlation matrix. We have a small performance drop when we project the optimal RIS phase-shift matrix to obtain unit-modulus entries. Nevertheless, using the optimized phase-shifts is advantageous over using DFT with the RS-LS method.

\begin{figure}[t!]
\hspace{0mm} 
\includegraphics[trim={0.7cm 0.1cm 1.5cm 1cm},clip,width=3.2in]{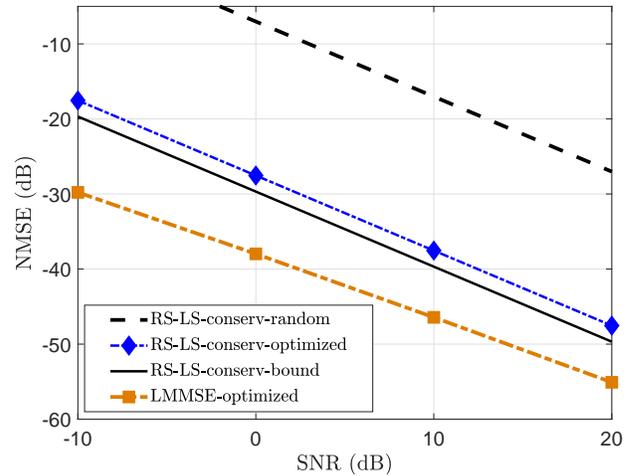}
			\vspace{-0.1cm}
			\caption{NMSE versus SNR for different estimators and RIS phase-shift configurations with $\tau_p=\overline{r}_{\rm RIS}+1=107$.} \label{fig:RSLS-taup-short} \vspace{-4mm}
\end{figure}

Fig.~\ref{fig:RSLS-taup-short} considers a reduced pilot length  $\tau_p=\overline{r}_{\rm RIS}+1$, where $\overline{r}_{\rm RIS}=106$ is the effective rank of $\overline{\vect{R}}_{\rm RIS}=\left(\vect{R}_{\rm RIS, iso}\odot\vect{R}_{\rm RIS, iso}\right)$ containing a fraction $1-10^{-5}$ of the sum of all eigenvalues.\footnote{We added one to the effective rank number $\overline{r}_{\rm RIS}$ to  prevent numerical issues due to having an ill-conditioned system.} Due to $\tau_p<N$ and the resulting rank deficiency in \eqref{eq:RS-LS-estimate}, the conventional LS estimator and the RS-LS estimator with DFT configuration cannot be used in the setup of Fig.~\ref{fig:RSLS-taup-short}. On the other hand, the proposed RS-LS estimator provides decent channel estimation accuracy when $\tau_p$ is less than half of $N$.  We also consider ``random'' phase-shifts as the alternative benchmark for which the unit-modulus entries have angles that are independently drawn from a uniform distribution on $[0,2\pi)$.  The gap between the randomized and optimized phase-shifts is more than 20\,dB, thus proving the effectiveness of the proposed optimized phase-shift method.

\section{Conclusions}
To overcome the channel estimation complexity issue in RIS-aided communications with a large number of antennas/elements, we proposed a novel estimator, called  RS-LS (reduced-space least-squares), that exploits only the array geometries and the resulting low-rank structure of any channel. This structure is not UE-specific or time-varying. Exploiting this structure allows us to outperform the LS estimator significantly. In addition, unlike the LS estimator, the RS-LS estimator can be utilized with much shorter pilot length. When the pilot length is small, optimizing the RIS configuration during training is necessary to obtain accurate  estimates.

\bibliographystyle{IEEEtran}
\bibliography{IEEEabrv,refs}

\end{document}